\documentclass{aa}  

\usepackage{graphicx}
\usepackage{txfonts}
\usepackage{natbib,twoopt}
\usepackage[breaklinks=true]{hyperref} 
\bibpunct{(}{)}{;}{a}{}{,}             
\makeatletter
  \newcommandtwoopt{\citeads}[3][][]{\href{http://adsabs.harvard.edu/abs/#3}%
    {\def\hyper@linkstart##1##2{}%
     \let\hyper@linkend\@empty\citealp[#1][#2]{#3}}}
  \newcommandtwoopt{\citepads}[3][][]{\href{http://adsabs.harvard.edu/abs/#3}%
    {\def\hyper@linkstart##1##2{}%
     \let\hyper@linkend\@empty\citep[#1][#2]{#3}}}
  \newcommandtwoopt{\citetads}[3][][]{\href{http://adsabs.harvard.edu/abs/#3}%
    {\def\hyper@linkstart##1##2{}%
     \let\hyper@linkend\@empty\citet[#1][#2]{#3}}}
  \newcommandtwoopt{\citeyearads}[3][][]%
    {\href{http://adsabs.harvard.edu/abs/#3}
    {\def\hyper@linkstart##1##2{}%
     \let\hyper@linkend\@empty\citeyear[#1][#2]{#3}}}
\makeatother

\begin{document}

   \title{NGC1333-IRAS4A as seen by the CALYPSO IRAM-PdBI survey\thanks{Based on observations carried out with the IRAM Plateau de Bure
Interferometer. IRAM is supported by INSU/CNRS (France), MPG (Germany), and IGN (Spain).}
   }

   \title{Jet multiplicity in the proto-binary system NGC1333-IRAS4A
   }

   \subtitle{The detailed CALYPSO IRAM-PdBI view\thanks{Based on observations carried out with the IRAM Plateau de Bure Interferometer. IRAM is supported by INSU/CNRS (France), MPG (Germany), and IGN (Spain).}}

   \author{G. Santangelo
          \inst{1,2}
          \and
          C. Codella
          \inst{1}
          \and
          S. Cabrit
          \inst{3,4,5,6}
          \and
          A.~J. Maury
          \inst{7}
          \and
          F. Gueth 
          \inst{8}
          \and
          S. Maret
          \inst{5,6}
          \and
          B. Lefloch
          \inst{5,6}
          \and
          A. Belloche
          \inst{9}
          \and
          Ph. Andr{\'e}
          \inst{7}
          \and
          P. Hennebelle
          \inst{7}
          \and
          S. Anderl
          \inst{5,6}
          \and
          L. Podio
          \inst{1}
          \and
          L. Testi
          \inst{10,1}
          }

   \institute{Osservatorio Astrofisico di Arcetri, Largo Enrico Fermi 5, 
              I-50125 Florence, Italy \\
              \email{gina.santangelo@oa-roma.inaf.it}
         \and
             Osservatorio Astronomico di Roma, via di Frascati 33, 00040 Monteporzio Catone, Italy
         \and             
             LERMA, Observatoire de Paris, PSL Research University, CNRS, UMR 8112, F-75014, Paris France
         \and             
             Sorbonne Universit{\'e}s, UPMC Univ. Paris 6, UMR 8112, LERMA, F-75005, Paris, France 
         \and
            Univ. Grenoble Alpes, IPAG, F-38000 Grenoble, France
         \and
            CNRS, IPAG, F-38000 Grenoble, France 
         \and
            Laboratoire AIM-Paris-Saclay, CEA/DSM/Irfu -- CNRS -- Universit{\'e} Paris Diderot, CE-Saclay, 91191 Gif-sur-Yvette, France
         \and
            IRAM, 300 rue de la Piscine, 38406 Saint Martin d'H{\`e}res, France
         \and
            Max-Planck-Institut f{\"u}r Radioastronomie, Auf dem H{\"u}gel 69, 53121 Bonn, Germany
         \and
            ESO, Karl-Schwarzschild-Strasse 2 D-85748 Garching bei M{\"u}nchen, Germany
             }

   \date{Received April 16, 2015; accepted August 17, 2015}

 
  \abstract
   {Owing to the paucity of sub-arcsecond (sub)mm observations required to probe the innermost regions of 
newly forming protostars, several fundamental questions are 
still being debated, such as the existence and coevality of close multiple systems. 
   }
   {
We study the physical and chemical properties of the jets and protostellar sources in 
the NGC1333-IRAS4A proto-binary system using continuum emission and molecular tracers of shocked gas.
   }
   {We observed NGC1333-IRAS4A in the SiO(6$-$5), SO(6$_5$$-$5$_4$), and CO(2$-$1) lines 
 and the continuum emission at 1.3, 1.4, and 3~mm using the IRAM Plateau de Bure Interferometer 
 in the framework of the CALYPSO large program.
  }
   {We clearly disentangle for the first time the outflow emission from the two sources A1 and A2. 
The two protostellar jets have very different properties: the A1 jet is faster, has a  
short dynamical timescale ($\lesssim$10$^3$ yr), 
and is associated with H$_2$ shocked emission, whereas the A2 jet, which dominates 
the large-scale emission, is associated with diffuse emission, bends, 
and emits at slower velocities.
The observed bending of the A2 jet is consistent with the change of propagation direction observed at large scale
and suggests jet precession on very short timescales ($\sim$200$-$600 yr).
In addition, a chemically rich spectrum with emission from several complex organic molecules 
(e.g. HCOOH, CH$_3$OCHO, CH$_3$OCH$_3$) is only detected towards A2.
Finally, very high-velocity shocked emission ($\sim$50 km s$^{-1}$) is observed along the A1 jet.  
An LTE analysis shows that SiO, SO, and H$_2$CO abundances in the gas phase are enhanced 
up to (3$-$4)$\times$10$^{-7}$, (1.4$-$1.7)$\times$10$^{-6}$, and (3$-$7.9)$\times$10$^{-7}$, respectively.
   }
   {The intrinsic different properties of the jets and driving sources in NGC1333-IRAS4A suggest 
different evolutionary stages for the two protostars, with A1 being younger than A2, in a very early stage of star formation 
previous to the hot-corino phase.
   }

               
   \titlerunning{Jet multiplicity in the proto-binary system NGC1333-IRAS4A}

   \maketitle
%

\section{Introduction}
\label{sec:intro}

Although multiple systems are a common outcome of the star formation process, the mechanism for their formation 
is still being debated \citep[e.g.][and references therein]{tohline2002,reipurth2014}. In this context, 
information about the system's multiplicity, coevality, and environment are particularly crucial. 
The ideal laboratories for these studies are very young sources that have not undergone significant evolution 
and are thus likely keep a memory of the initial physical and chemical conditions.
Class 0 protostars, with lifetimes of less than a few 10$^5$ yr, represent 
the earliest stage of low-mass star formation, when most of the mass is still 
in the form of a dense infalling envelope \citep[e.g.][]{andre2000,evans2009}.  
Once it is sufficiently luminous, the central protostar radiatively heats the surrounding inner envelope, and
grain mantle evaporation occurs, which triggers a hot-corino chemistry rich in complex organic molecules (COMs)
\citep[][]{vandishoeck1998,ceccarelli2004,bottinelli2004}. 
Protostars also drive fast jets \citep[e.g.][]{arce2007,ray2007,frank2014} that impact the high-density parent cloud and generate shock fronts. 
This leads to significant enhancements in the abundance 
of several molecules, such as H$_2$O, CH$_3$OH, and SiO \citep[e.g.][]{vandishoeck1998}.
High angular resolution studies of Class 0 protostars are thus needed to probe the innermost regions ($\leq$100 AU) 
with the aim of disentangling the emission originating from the different processes: 
the launch of the jet, the outflow cavity emission, and the chemistry of the hot corino. 
We present here new observations of molecular tracers of shocked gas 
and continuum emission towards the NGC1333-IRAS4A protostellar system
(hereafter IRAS4A) located at 235 pc \citep{hirota2008}. 
Observations are part of the Continuum and Lines in Young ProtoStellar Objects 
(CALYPSO)\footnote{http://irfu.cea.fr/Projects/Calypso} large program, 
aimed at studying a large sample of Class 0 protostars at sub-arcsecond resolution 
with the IRAM Plateau de Bure Interferometer (hereafter PdBI).

IRAS4A has been identified as a binary Class 0 system using mm interferometry 
\citep[][]{jennings1987,sandell1991,lay1995,looney2000}. 
Since the binary components have never been resolved in the
infrared band, the luminosity of each protostar is not known. 
The bolometric luminosity of the IRAS4A system has been derived by \citet{enoch2009} to be $\sim$4.2~$L_{\sun}$
and later by \citet{kristensen2012} and \citet{karska2013} to be $\sim$9.1~$L_{\sun}$ using PACS continuum measurements.
The two Class 0 components, IRAS4A1 and IRAS4A2 (hereafter A1 and A2), 
have a separation of only about 1.8$^{\prime\prime}$ ($\sim$420 AU). 
A1 is more than three times brighter in the millimetre and centimetre continuum
than its companion A2 \citep{looney2000,reipurth2002,joergensen2007}, 
whereas it is weaker in NH$_3$ emission \citep{choi2007}. 
The IRAS4A system is associated with a spectacular large-scale (a few arcminutes) 
bipolar outflow (PA$\sim$45$^{\circ}$) observed in several tracers, such as 
CO, SiO, and H$_2$O \citep[e.g.][]{blake1995,choi2001,choi2005,yildiz2012,santangelo2014}.
A shorter southern monopolar blue-shifted lobe 
(PA$\sim$$-$10$^{\circ}$) is observed in SiO by \citet{choi2005} with no clear northern counterpart. 
However, none of the observations to date resolve the emission within $\sim$10$^{\prime\prime}$ from the 
protostars and therefore they do not allow us to disentangle the outflow emission 
and directly identify the driving sources.
In this context, the new high angular resolution observations of CO, SiO, and SO presented here 
allow us to probe the innermost region of IRAS4A. 
The goal is to study jet multiplicity and the physical and chemical properties of the jet and the driving sources.


\section{Observations}
\label{sec:obs}

IRAS4A was observed at 1.3, 1.4, and 3~mm with the IRAM PdBI between February 2011 
and February 2013 using both the A and C configurations. 
The baseline range of the observations is 15$-$762 m, 
which allows us to recover emission on scales from 
18$^{\prime\prime}$ to 0$.\!\!^{\prime\prime}$35 at 1.3 mm.
WideX backends were used to cover the full 3.8 GHz spectral window at the spectral resolution 
of 2 MHz ($\sim$2.7 km s$^{-1}$ at 1.4 mm). 
In addition, higher resolution backends were employed to observe the 
SiO(5$-$4), SO(6$_5$$-$5$_4$), and CO(2$-$1) lines at 217.105, 219.949, and 230.538 GHz.
Calibration was carried out following standard procedures,
using GILDAS-CLIC\footnote{http://www.iram.fr/IRAMFR/GILDAS/}.
The phase RMS is $\leq$50$^{\circ}$ and $\leq$80$^{\circ}$
for the A and C tracks, with precipitable water vapour (PWV) between 0.5 mm and 2 mm 
and system temperatures between 100 K and 250 K. 
The uncertainty on the absolute flux scale is $\sim$10\%. 
The synthesised FWHM beam is 0$.\!\!^{\prime\prime}$7$\times$0$.\!\!^{\prime\prime}$4 (PA = 19$^{\circ}$) at 1.3 mm, 
1$.\!\!^{\prime\prime}$1$\times$0$.\!\!^{\prime\prime}$8 (19$^{\circ}$) at 1.4 mm,
and 1$.\!\!^{\prime\prime}$8$\times$1$.\!\!^{\prime\prime}$3 (39$^{\circ}$) at 3 mm.
The typical RMS noise per 1~km s$^{-1}$ channel is 
$\sim$8 mJy/beam at 1.3 mm, $\sim$10 mJy/beam at 1.4 mm, and $\sim$3 mJy/beam at 3 mm.

%

\section{Results}
\label{sec:results}

\subsection{Continuum emission: source multiplicity and properties}
\label{subsec:continuum}

\begin{figure}
\centering
\includegraphics[width=0.49\textwidth]{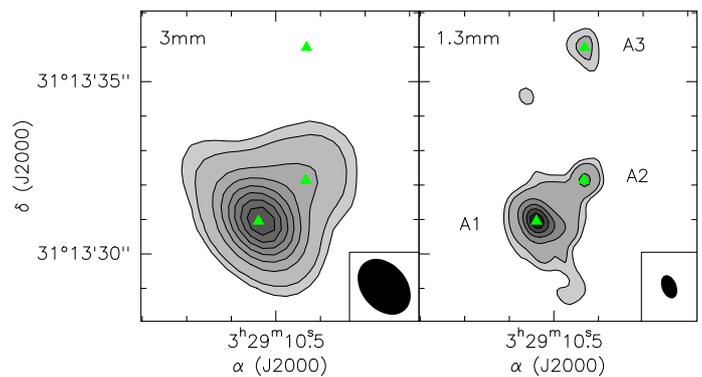}
\caption{\emph{Left:} 3~mm continuum emission map of the IRAS4A region (grey scale and black contours). 
The contour levels of emission are traced at 5 and 8~$\sigma$ levels and increase in steps of 10~$\sigma$, 
where $\sigma$$=$2~mJy~beam$^{-1}$. The positions of the continuum sources at 1.3~mm are marked in green.
The PdBI synthesised beam (HPBW) of the continuum map is shown. 
\emph{Right:} 1.3~mm continuum map of IRAS4A. Contour levels and labels are the same as in the left panel. 
The rms noise is $\sigma$$=$8~mJy~beam$^{-1}$.
}
\label{fig:continuum}
\end{figure}

\begin{table}
\caption{Parameters of the continuum peaks.}
\label{table:continuum}
\centering
\begin{tabular}{l c c c c }
\hline\hline
&&& \multicolumn{2}{c}{Flux\tablefootmark{a}}  \\
\cline{4-5}
Source & $\alpha$(J2000) & $\delta$(J2000) & 1.3mm & 3mm \\ 
& (03$^{\rm h}$ 29$^{\rm m}$) & (31$^{\circ}$ 13$^{\prime}$) & (mJy) & (mJy) \\
\hline
A1 & 10$.\!\!^{\rm s}$53 & 31$.\!\!^{\prime\prime}$0 &   1230   &     144     \\  
A2 & 10$.\!\!^{\rm s}$42 & 32$.\!\!^{\prime\prime}$3 &     369   &       40      \\   
A3 & 10$.\!\!^{\rm s}$44 & 35$.\!\!^{\prime\prime}$9 &     187   &   $<$2      \\
\hline
\end{tabular}
\tablefoot{
\tablefoottext{a}{Fluxes are measured by integrating over 1$^{\prime\prime}$. The uncertainty on the absolute flux is 10\%, which translates into an uncertainty of 15\% on the spectral indexes (see Sect.~\ref{subsec:continuum}).}
}
\end{table}

Emission maps of the 1.3 and 3~mm continuum are shown in Fig.~\ref{fig:continuum}. 
Both known protostars A1 and A2 are detected in the continuum emission at all frequencies with A1 being brighter than A2.
They are well resolved at 1.3~mm, with a separation of $\sim$1$.\!\!^{\prime\prime}$8. 
In addition, a third continuum source (hereafter A3) is detected for the first time only at 1.3~mm at a position 
($+$0$.\!\!^{\prime\prime}$2, $+$3$.\!\!^{\prime\prime}$7) offset from A2. 
Positions and integrated fluxes of the three continuum sources are derived
by performing a power-law fit to the continuum visibilities of the detected sources (see Table~\ref{table:continuum}).

We estimate the spectral index of the continuum emission $\alpha_{\rm 3~mm}^{\rm 1.3~mm}$ 
(where flux density is $F_{\nu} \propto \nu^{\alpha}$) for the three detected sources. 
Spectral indexes of 2.3 and 2.4 are obtained for sources A1 and A2, respectively, which are consistent with 
dust thermal radiation from 
embedded protostellar objects.
On the other hand, the non-detection of source A3 at 1.4 and 3 mm implies a steep spectrum for this source with a spectral index 
$\alpha_{\rm 3~mm}^{\rm 1.3~mm}$$>$4.9 
(based on the upper limit on the 3~mm flux density), which is not consistent with a protostellar nature. 
Compact continuum emission along protostellar jets has already been observed in other sources 
\citep[e.g.][]{gueth2003,maury2010,codella2014}; envelope emission or shock-heated dust along the jet
have been suggested as possible explanations for the origin of these features. 
However, a larger statistically suitable sample is crucial in order to assess this issue. 
A detailed analysis of the continuum emission in the whole CALYPSO sample 
will be presented in a forthcoming paper.

\subsubsection{Hot corino at A2}
\label{subsubsec:hotcorino}

\begin{figure}[h!]
\centering
\includegraphics[width=0.45\textwidth]{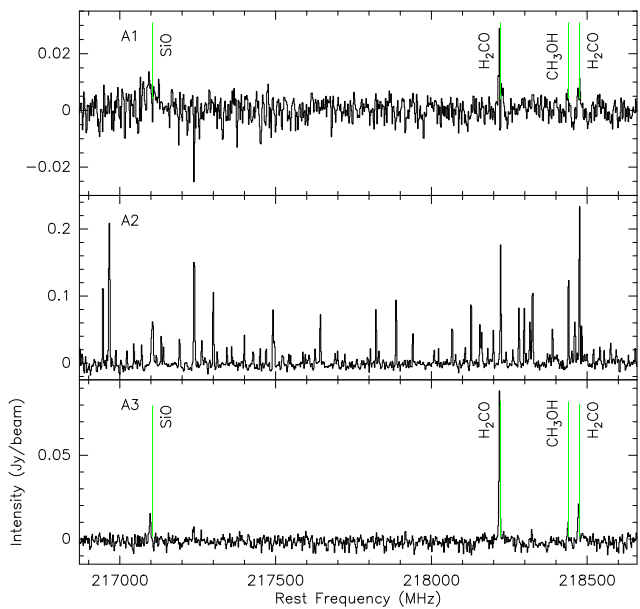}
\caption{WideX spectra at 1.4~mm at the position of A1 (\emph{upper panel}), A2 (\emph{middle panel}), 
and A3 (\emph{lower panel}). 
We note the typical hot-corino chemically rich spectrum observed towards A2. As already shown by \citet{taquet2015}, 
the spectrum presents emission from several COMs, such as CH$_3$OH (methanol), HCOOH (formic acid), 
CH$_3$OCHO (methyl formate), and CH$_3$OCH$_3$ (dimethyl ether).
}
\label{fig:widex_A123}
\end{figure}

Figure~\ref{fig:widex_A123} presents the WideX spectra observed at the three continuum source positions. 
A chemically rich spectrum is detected only towards A2, highlighting a hot-corino chemistry for this protostar.
As already shown by \citet{taquet2015}, emission from several COMs is indeed observed at A2, 
such as CH$_3$OH (methanol), HCOOH (formic acid), CH$_3$OCHO (methyl formate), 
and CH$_3$OCH$_3$ (dimethyl ether); similar molecules are observed towards the Class 0 protostar IRAS2A \citep{maury2014}.
A detailed identification and analysis of the detected lines 
will be presented in a forthcoming paper. For the scope of this paper, this emission is indicative of a rich hot-corino 
chemistry associated with source A2.
On the other hand, only SiO, H$_2$CO, and CH$_3$OH are clearly detected towards the continuum sources A1 and A3.

\subsubsection{Infall signatures}
\label{subsubsec:infall}

\begin{figure}[h!]
\centering
\includegraphics[width=0.45\textwidth]{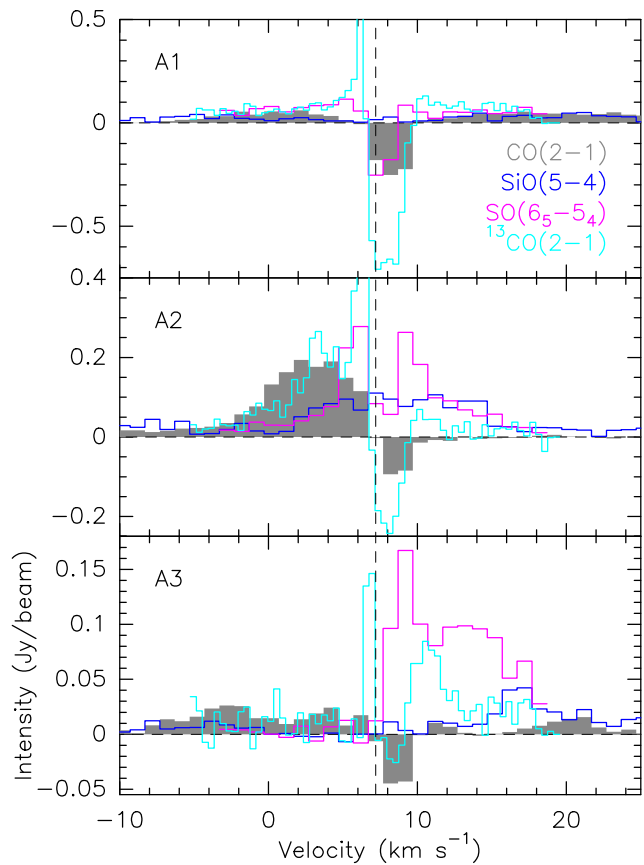}
\caption{$^{12}$CO(2$-$1) (grey), SiO(5$-$4) (blue), SO(6$_5$$-$5$_4$) (magenta), and $^{13}$CO(2$-$1) (cyan)
at the positions of the continuum sources A1, A2, and A3 (see Fig.~\ref{fig:channelmaps}). 
The selected positions are displayed from top to bottom. 
The $^{12}$CO(2$-$1) data have been convolved to match the angular resolution at 1.4~mm 
(1$.\!\!^{\prime\prime}$1$\times$0$.\!\!^{\prime\prime}$8).
The vertical dashed line marks the systemic velocity of $+$7.2~km~s$^{-1}$ for the 
three continuum sources. We note that the systemic velocity of A3 
has been derived from PdBI N$_2$H$^+$(1$-$0) observations at 93.17 GHz from our CALYPSO dataset.
}
\label{fig:spettri_A123}
\end{figure}

Figure~\ref{fig:spettri_A123} presents the high-resolution spectra at the continuum source positions of the selected 
molecular species. Absorption is detected 
towards A1 and A2 in $^{12}$CO, $^{13}$CO, and SO at similar slightly red-shifted velocities ($\sim$8.2~km s$^{-1}$).
In particular, the $^{13}$CO spectra show inverse P-Cygni profiles, which are indicative of infall motions.
Similar profiles were previously detected in H$_2$CO and CS 
with PdBI and NRO-NMA by \citet{difrancesco2001}. The authors modelled the emission with a two-layer radiative transfer code
and interpreted it as infall in the warm gas associated with the young stellar objects. 
Inverse P-Cygni profiles were also observed in $^{13}$CO with the SMA by \citet{joergensen2007} 
and interpreted as larger scale infall motions in the ambient cloud or outer envelope rather than collapse 
onto the central protostars.
For the first time, we spatially resolve the absorption and detect infall signatures associated with both 
known protostars A1 and A2 at similar velocity. Whether the detected spectra presented here probe infall towards each source 
or global infall of a common envelope is not clear from our data and a detailed modelling of the line profiles is certainly 
needed in order to resolve this issue. However, this is beyond the scope of this paper.
We note that infall motions on larger scales associated with the envelope of IRAS4A 
were observed in HCN by \citet{choi1999} and in CS, C$^{34}$S, and N$_2$H$^+$ by \citet{belloche2006}.

Interestingly, red-shifted absorption is also detected in $^{12}$CO and $^{13}$CO towards the newly detected 
continuum source A3. 

\subsection{Line emission: jet multiplicity}
\label{subsec:jets-multi}

\begin{figure*}
\centering
\includegraphics[width=0.65\textwidth]{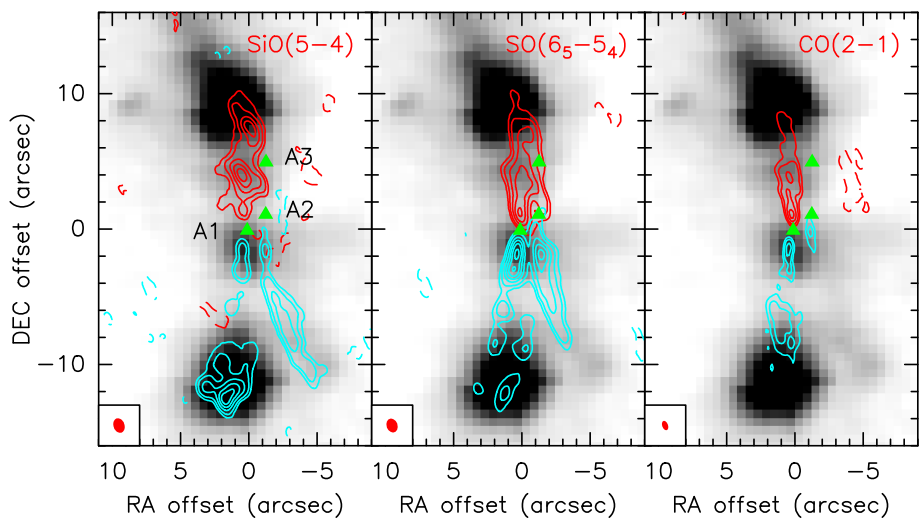}
\caption{\emph{Left:} The \emph{Spitzer}-IRAC map of band~2 at 4.5~$\mu$m (grey scale) 
is compared with the SiO(5$-$4) map integrated in the blue- (-20, +5~km~s$^{-1}$, cyan contours) 
and red-shifted (+10, +55~km~s$^{-1}$, red contours) velocity ranges.
Offsets are with respect to IRAS4A1. 
The contour levels of emission are traced at 5 and 8~$\sigma$ levels and increase in steps of 5~$\sigma$, 
where $\sigma$$=$100~mJy~beam$^{-1}$ for the red-shifted emission and 70~mJy~beam$^{-1}$ for the blue-shifted emission.
Negative emission is shown with dashed contours starting at the 3~$\sigma$ level and decreasing in steps of 3~$\sigma$.
The positions of the continuum sources at 1.3~mm are marked in green. 
The PdBI synthesised beam (HPBW) is shown in the bottom-left corner.
\emph{Middle} and \emph{Right:} Same as the left panel for the SO(6$_5$$-$5$_4$) and $^{12}$CO(2$-$1) emissions, respectively.
The rms noise of the velocity-integrated maps are: $\sigma$$=$110~mJy~beam$^{-1}$ and 90~mJy~beam$^{-1}$ for the red- 
and blue-shifted SO(6$_5$$-$5$_4$) emissions, respectively; and $\sigma$$=$500~mJy~beam$^{-1}$ and 400~mJy~beam$^{-1}$
for the red- and blue-shifted $^{12}$CO(2$-$1) emissions, respectively.
}
\label{fig:h2_sio_so_co}
\end{figure*}

\begin{figure*}
\centering
\includegraphics[width=0.75\textwidth]{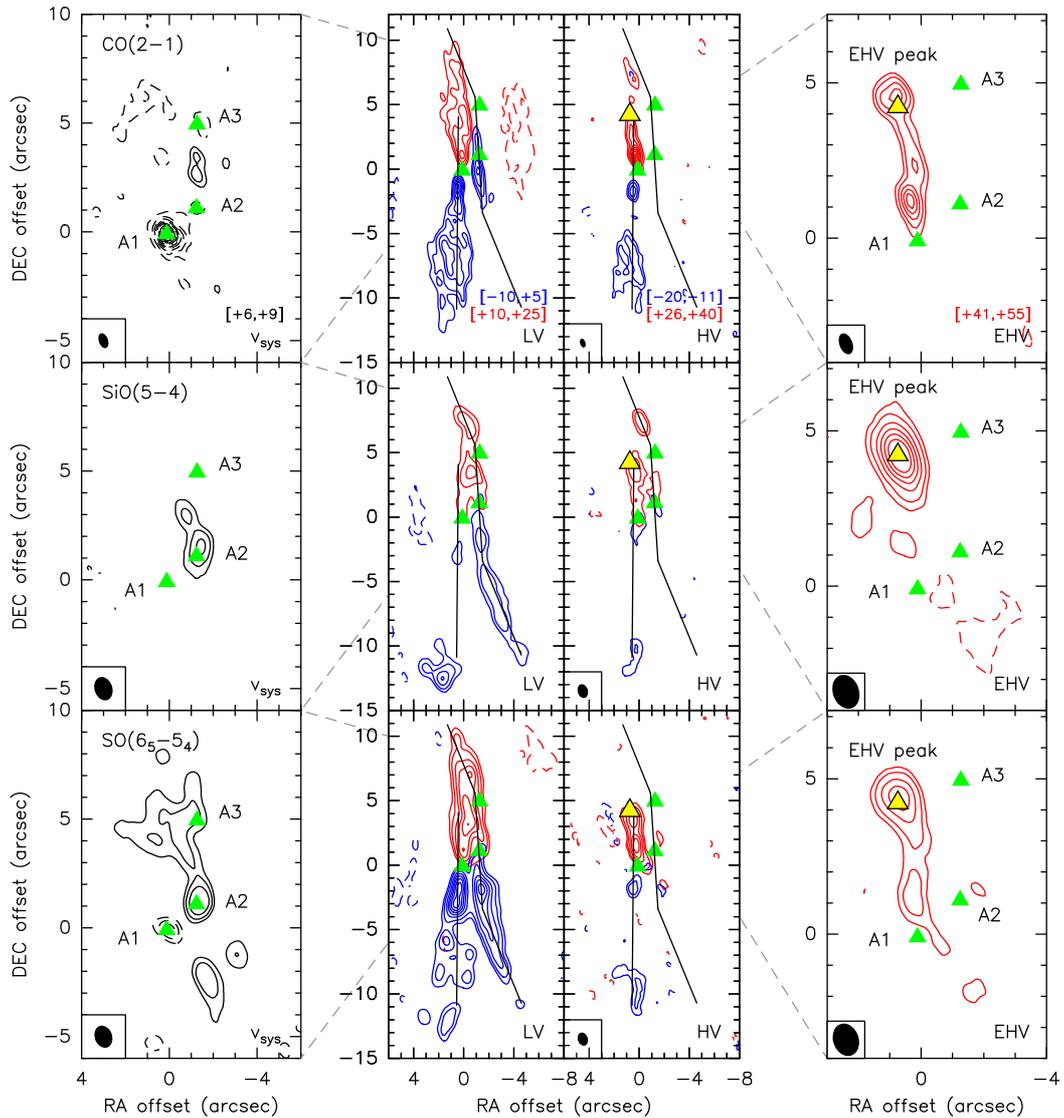}
\caption{\emph{Upper:} Channel maps of the $^{12}$CO(2$-$1) emission integrated in the ambient systemic- ($v_{\rm sys}$; +6, +9~km~s$^{-1}$), 
low- (LV; -10, +5~km~s$^{-1}$ and +10, +25~km~s$^{-1}$), high- (HV; -20, -11~km~s$^{-1}$ and +26, +40~km~s$^{-1}$), 
and extremely high- (EHV, +41, +55~km~s$^{-1}$) velocity ranges.
The contour levels of emission are traced at 5 and 8~$\sigma$ levels and increase in steps of 5~$\sigma$. 
Negative emission is shown with dashed contours starting at the 3~$\sigma$ level and decreasing in steps of 3~$\sigma$.
The positions of the continuum sources at 1.3~mm are marked in green and the position of the SiO(5$-$4) emission peak 
in the EHV range is marked in yellow. The PdBI synthesised beams (HPBW) are shown in the bottom-left corner. 
Solid black curves indicate the proposed propagation directions of the A1 and A2 jets.
\emph{Middle:} Same as the upper panel for the SiO(5$-$4) emission.
\emph{Lower:} Same as the upper panel for the SO(6$_5$$-$5$_4$) emission.
}
\label{fig:channelmaps}
\end{figure*}

Figure~\ref{fig:h2_sio_so_co} shows the velocity-integrated emission of 
SiO(5$-$4), SO(6$_5$$-$5$_4$), and $^{12}$CO(2$-$1) towards IRAS4A, 
in comparison with the \emph{Spitzer}-IRAC map of band~2 at 4.5~$\mu$m. 
This IRAC band covers several H$_2$ emission lines, which are particularly strong in shocked regions 
making IRAC images good tracers of H$_2$ emission \citep[e.g.][]{reach2006,neufeld2008}.
For the first time the bipolar outflow is clearly disentangled in the southern blue-shifted emission 
into two different flows tracing the jets driven by the A1 and A2 protostars. 
SiO emission is detected for the first time in IRAS4A 
at close distance from the continuum sources, 
inside $\sim$10$^{\prime\prime}$ \citep[$\sim$2400 AU, see][]{choi2005,choi2006}.
Strong IRAC H$_2$ emission is only associated with the jet from A1, 
which possibly indicates a stronger shock or the presence of denser ambient material.
In particular, the bright southern and northern H$_2$ knots in the IRAC map coincide with the 
tips of the blue-shifted and red-shifted (see also Fig.~\ref{fig:channelmaps}) A1 lobes seen in our SiO map. 
Since there is no H$_2$ or single-dish CO outflow signature further along the A1 jet axis, as can be seen on larger scales 
from the IRAC and APEX maps in \citet{santangelo2014}, it appears that here we are tracing the terminal shocks, 
where the A1 outflow impacts the ambient cloud.
Remarkably, our observations detect a sharp bend to the south-west of the blue-shifted A2 jet  
at a distance of about 4$^{\prime\prime}$ from the driving source. On the other hand, emission from 
the two sources cannot be disentangled in the velocity-integrated SiO red-shifted northern emission, possibly because 
geometric effects cause mixing of the emission; possible additional confusion may be due to the 
presence of the continuum source A3.

The two jets are also unraveled in the SO emission. The SO morphology is different 
with respect to SiO, with the latter peaking at the tip of the blue-shifted lobe of the A1 outflow, in association with the IRAC H$_2$ emission, 
and clearly delineating the A2 jet over more than 10$^{\prime\prime}$ all the way down to the source, whereas the SO emission
peaks closer to the protostars.
Finally, in the CO emission only the A1 jet is clearly detected; the strongest peak of the blue-shifted lobe 
appears to be shifted with respect to the IRAC H$_2$ and SiO emissions, further suggesting a chemical 
differentiation in the jet.
Faint and compact blue-shifted CO emission associated with the A2 jet is only detected close to the driving source, before the bend.

\subsection{Line emission: jet kinematics}
\label{subsec:jets-kinematics}

A more detailed view of the region, showing in particular the multiplicity and differentiation of the A1 and A2 jets, 
is given by the channel maps of the molecular emission presented in Fig.~\ref{fig:channelmaps}. 
Four velocity ranges are identified in CO, SiO, and SO emissions, corresponding to 
the systemic ($v_{\rm sys}$$\pm$1.5~km~s$^{-1}$), low-velocity 
(LV, from $v_{\rm sys}$$\pm$2.5~km~s$^{-1}$ to $v_{\rm sys}$$\pm$17.5~km~s$^{-1}$), 
high-velocity (HV, from $v_{\rm sys}$$+$18.5~km~s$^{-1}$ to $v_{\rm sys}$$+$32.5~km~s$^{-1}$ and from 
$v_{\rm sys}$$-$18.5~km~s$^{-1}$ to $v_{\rm sys}$$-$27.5~km~s$^{-1}$), 
and extremely high-velocity (EHV, from $v_{\rm sys}$$+$33.5~km~s$^{-1}$ to $v_{\rm sys}$$+$47.5~km~s$^{-1}$) gas.
Compact and spatially unresolved absorption is detected around systemic velocity 
\citep[$v_{\rm sys}$$=$$+$7.2~km~s$^{-1}$,][]{belloche2006} 
towards A1, A2, and A3 in CO emission and only towards A1 in SO,
as already shown in Sect.~\ref{subsubsec:infall}.
An extended structure with A2 at the vertex is seen in SO emission at the systemic velocity, 
possibly probing the hot corino and outflow cavity walls 
\citep[see also Sect.~\ref{subsubsec:hotcorino} and][for NGC1333-IRAS2A]{codella2014}. 

The molecular emissions integrated over the higher velocity channels show that 
we are finally able to disentangle both the blue- and red-shifted outflow emissions 
associated with the two protostars A1 and A2. In particular, we resolve  
the red-shifted counterpart of the A1 southern outflow for the first time.
The A1 jet appears to be faster than the A2 jet 
as shown by its association with HV and EHV emissions; 
in particular, jet-like gas at red-shifted EHV is detected in association with the A1 jet, showing a strong compact peak about 
4$^{\prime\prime}$ north with respect to A1.
On the other hand, the A2 jet is brightest in the LV range, with only faint molecular line emission in the HV range.
The sharp bend of the A2 jet observed at these small scales in the velocity-integrated blue-shifted 
SiO and SO emissions (see previous section) is detected in the LV and HV channel maps. 
They also clearly show a symmetric bending in the northern red-shifted A2 jet emission at about 4$^{\prime\prime}$ from A2, 
which is close to the continuum source A3.
The symmetric morphology of the bend suggests a jet precession on very short timescales.  
Indeed, assuming an observed velocity of 10$-$20~km~s$^{-1}$ (see Fig.~\ref{fig:spettri_A123})
and an inclination with respect to the line of sight of about 45$^{\circ}$$-$60$^{\circ}$ \citep[see][]{yildiz2012}, 
the dynamical age of the bend is about 200$-$600 yr.
After this bending, the axes of the A2 jet correspond to the axes of the large-scale SiO and CO outflow emissions  
\citep[see e.g.][and Fig.~\ref{Afig:SD} in Appendix~\ref{Asec:SD}]{blake1995, girart1999,choi2001,choi2005}, confirming that 
the large-scale outflow emission is driven by the A2 protostar.
In this context, the non-detection of CO(2$-$1) emission associated with the A2 jet 
at distance larger than $\sim$3$^{\prime\prime}$ even in the LV range may be due to the 
filtering out of extended emission by the interferometer, given the high angular resolution at 1.3~mm.

\section{Analysis and discussion}
\label{sec:discussion}

\subsection{Chemical abundances in the A1 jet at the EHV peak}
\label{subsec:EHVgas}

\begin{figure}
\centering
\includegraphics[width=0.45\textwidth]{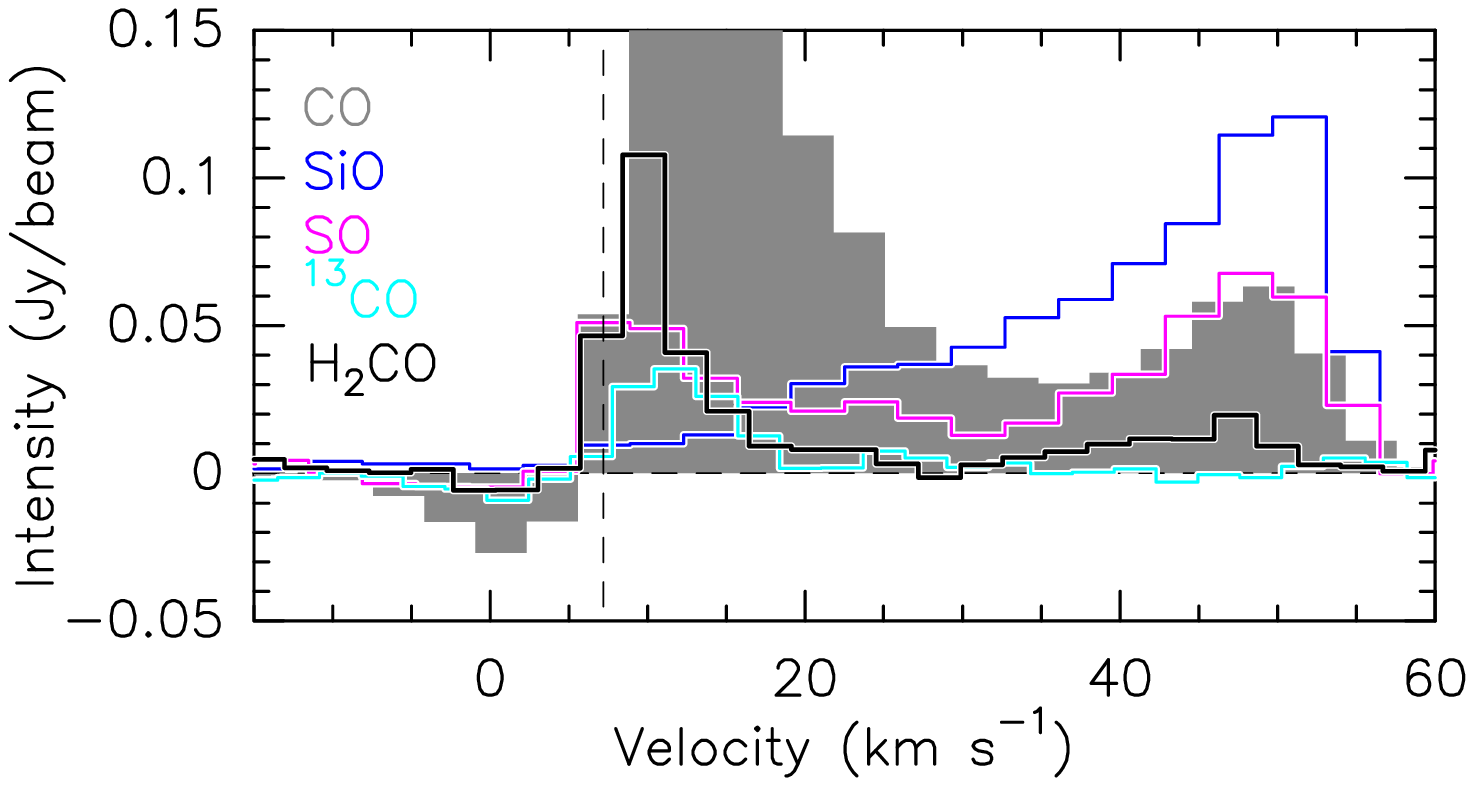}
\caption{$^{12}$CO(2$-$1) (grey), SiO(5$-$4) (blue), SO(6$_5$$-$5$_4$) (magenta), $^{13}$CO(2$-$1) (cyan), 
and p-H$_2$CO(3$_{03}$$-$2$_{02}$) (black) spectra
at the EHV peak (see Fig.~\ref{fig:channelmaps}). The p-H$_2$CO(3$_{03}$$-$2$_{02}$) spectrum 
was observed with WideX (see Appendix~\ref{Asec:widex}) at the spectral resolution of $\sim$2.7~km~s$^{-1}$;  
all other spectra have thus been smoothed to the same spectral resolution.
The $^{12}$CO(2$-$1) data have been convolved to match the angular resolution at 1.4~mm (1$.\!\!^{\prime\prime}$1$\times$0$.\!\!^{\prime\prime}$8).
}
\label{fig:spettri_EHV}
\end{figure}

\begin{table*}
\caption{Abundances with respect to H$_2$ of the observed species at 40 and 300~K for the LV and EHV gas.}
\label{table:abundances}
\centering
\begin{tabular}{l | c c c c c | c c c c c }
\hline\hline
& \multicolumn{5}{c|}{LV (9$-$25 km s$^{-1}$)} &  \multicolumn{5}{c}{(E)HV (26$-$55 km s$^{-1}$)} \\
& & \multicolumn{2}{c}{$T$=40~K} & \multicolumn{2}{c|}{$T$=300~K} & & \multicolumn{2}{c}{$T$=40~K} & \multicolumn{2}{c}{$T$=300~K} \\
Mol. & $\int Tdv$ & $N$ & $X$\tablefootmark{a} & $N$ & $X$\tablefootmark{a} & $\int Tdv$ & $N$ & $X$\tablefootmark{b} & $N$ & $X$\tablefootmark{b} \\
& (K~km~s$^{-1}$) & (cm$^{-2}$) & (10$^{-8}$) & (cm$^{-2}$) & (10$^{-8}$) & (K~km~s$^{-1}$) & (cm$^{-2}$) & (10$^{-8}$) & (cm$^{-2}$) & (10$^{-8}$) \\
\hline
$^{13}$CO   &  8.6  & 1.2$\times$10$^{16}$ &   --  & 6.4$\times$10$^{16}$ &   --   &  --     &               --                        &  --  &               --                        &  --  \\
$^{12}$CO\tablefootmark{c}   & --  &    --        &   --   &               --                       &    --   & 31.4 & 2.0$\times$10$^{16}$ &  -- & 1.1$\times$10$^{17}$ &  --  \\
SiO                &  --     &               --                        &   --   &               --                        &   --  & 61.9 & 8.3$\times$10$^{13}$ & 41 & 3.2$\times$10$^{14}$ & 29 \\
SO                 & 14.4 & 1.8$\times$10$^{14}$ & 1.9 & 7.3$\times$10$^{14}$ & 1.5 & 28.7 & 3.5$\times$10$^{14}$ & 170 & 1.5$\times$10$^{15}$ & 140 \\
p-H$_2$CO & 14.6 & 1.5$\times$10$^{14}$ & 1.7 & 2.1$\times$10$^{15}$ & 4.3 &  5.8 & 6.2$\times$10$^{13}$ &  30 & 8.5$\times$10$^{14}$ & 79 \\
\hline
\end{tabular}
\tablefoot{
\tablefoottext{a}{$X$($^{12}$C/$^{13}$C)$=$77 \citep{wilson1994}.}
\tablefoottext{b}{$X$($^{12}$CO/H$_2$)$=$$10^{-4}$.}
\tablefoottext{c}{The $^{12}$CO(2$-$1) data have been convolved to match the angular resolution at 1.4~mm (1$.\!\!^{\prime\prime}$1$\times$0$.\!\!^{\prime\prime}$8).}
}
\end{table*}

Red-shifted EHV (41$-$55~km~s$^{-1}$) emission is detected 
in CO, SiO, SO, and H$_2$CO
(see Figs.~\ref{fig:channelmaps}, \ref{fig:spettri_EHV}, and \ref{Afig:widexEHV}). 
It shows a compact ($\sim$1$.\!\!^{\prime\prime}$6$\times$1$.\!\!^{\prime\prime}$1 from visibility fitting) morphology in the SiO emission 
with the emission peak located ($+$0$.\!\!^{\prime\prime}$7, $+$4$.\!\!^{\prime\prime}$3) offset from A1, 
whereas a more elongated and collimated structure pointing to A1 is detected in the CO and SO emissions. 

A line profile comparison at the EHV peak reveals that 
CO, SO, and H$_2$CO show an emission peak at low velocity ($\sim$10~km~s$^{-1}$)
and a secondary EHV peak around $+$50~km~s$^{-1}$; the non-detection 
of EHV $^{13}$CO is due to a lack of sensitivity (if CO is optically thin, the expected $^{13}$CO at the EHV peak of $^{12}$CO
is approximately the same as the rms noise). 
On the other hand, the SiO line profile presents an emission peak 
at $+$50~km~s$^{-1}$ with a wing extending down to the systemic velocity. 
We note that this is the first detection of SiO emission at such high velocity 
in this source \citep[][]{choi2005,choi2011}.
The similarity between peak velocities and line profiles of the molecules detected at EHV suggests that 
they share a common origin, tracing the primary molecular jet \citep[see also][]{codella2014}.

Abundances of the detected species have been derived under the assumption of LTE conditions, 
using CO as a proxy for H$_2$ emission.
The emission has been divided into two velocity ranges: LV (9$-$25 km s$^{-1}$) and (E)HV (26$-$55 km s$^{-1}$).
For the analysis, two values of temperature have been considered: 40 and 300 K. 
Since $^{12}$CO emission is optically thick at velocities close to systemic ($\tau^{\rm ^{12}CO}$$\sim$11), 
we use $^{13}$CO ($\tau^{\rm ^{13}CO}$$\sim$0.1) to derive abundances in the LV range, 
assuming optically thin conditions and $^{12}$C/$^{13}$C$=$77 \citep{wilson1994}.
On the other hand, abundances in the (E)HV component were obtained using $^{12}$CO ($\tau^{\rm ^{12}CO}$$<$14),
assuming a typical CO abundance of 10$^{-4}$. 
Abundance enhancements are derived for all species in the high-velocity gas (Table~\ref{table:abundances}) 
and suggest a shock origin.
In particular, a SiO abundance of (2.9$-$4.1)$\times$10$^{-7}$ is derived in the high-velocity gas, 
consistent with previous single-dish estimates in Class 0 outflows 
\citep[e.g.][]{bachiller1997,tafalla2010} and in agreement with current shock model predictions
\citep{gusdorf2008a,gusdorf2008b,guillet2009,guillet2011}.
However, we note that \citet{cabrit2012} showed that PdBI SiO(5$-$4) emission in the HH212 protostellar jet 
can be optically thick at high velocities and derived an SiO abundance up to about 3$\times$10$^{-6}$. 

It has been observed that the SO abundance is enhanced by two orders of magnitude
from (1.5$-$1.9)$\times$10$^{-8}$ in the low-velocity gas to (1.4$-$1.7)$\times$10$^{-6}$ in the high-velocity gas. 
The derived abundances are 
slightly higher than the values observed by \citet{bachiller1997} and \citet{tafalla2010}, i.e. $\sim$(2$-$3)$\times$10$^{-7}$, 
although their estimates are based on single-dish observations.
On the other hand, an SO abundance of 2$\times$10$^{-6}$ is derived by \citet{lee2010} in the HH211 Class 0 jet
from SMA interferometric observations. 
\citet{pineaudesforets1993} and \citet{flower1994}
modelled SO emission in C-type shocks and predicted an increase of SO abundance in shock regions 
with maximum values of a few 10$^{-7}$, consistent with our estimates. 

Finally, an enhancement of the p-H$_2$CO abundance from (1.7$-$4.3)$\times$10$^{-8}$ 
to (3$-$7.9)$\times$10$^{-7}$ is observed between the low- and high-velocity gas components. 
This is not in agreement with the estimates for Class 0 outflows by \citet{tafalla2010}, 
who reported higher abundance values in the low-velocity gas 
(a few 10$^{-8}$ with respect to $<$10$^{-8}$ for the high-velocity gas). 
Their results, however, are based on single-dish observations.
It is worth noting that this molecule is expected to be released from icy mantles, 
which would support our detection of an abundance increase in fast shocks.
Given the size of the beam, the estimated H$_2$CO abundances are in agreement with C-type shock models 
including grain-grain processing, as described in \citet{anderl2013}.

\subsection{Different properties of the A1 and A2 jets}
\label{subsec:jet_differences}

The new PdBI data presented in this paper highlight the intrinsic differences between 
the jets driven by the two protostars A1 and A2. 
In particular, the jets present different morphologies with the A2 jet bending on small scales in a 
mirror-symmetric S-shaped pattern, while the A1 jet is C-shaped with 
the blue- and red-shifted emissions both tilted to the east of A1.
The A1 jet appears to be faster than the A2 jet, as can be seen from its association with 
bright HV and EHV emissions, whereas the A2 jet is mainly detected in the LV range.
Finally, the A2 jet shows a large spatial extent on the sky of more than 4$^{\prime}$, while 
the A1 jet is only about 20$^{\prime\prime}$ long. 
Three possible explanations may be invoked in order to account for the 
different observational characteristics of the jets associated with the two protostars: 
1) different inclinations for the jets, 2) different masses for the driving sources, or 3) different ages for jets and driving sources.

A jet inclination effect could explain the 
large difference in projected lengths for the two outflows. In this case, assuming the same intrinsic length 
for both jets (about 120$^{\prime\prime}$ for each lobe), the projected length of the A1 jet (about 13$^{\prime\prime}$ 
in the blue-shifted lobe) suggests that it should be viewed at about 6$^{\circ}$ from pole-on. 
This is, however, statistically very unlikely and not consistent with the morphology of the emission we observed 
\citep[see][for a review]{cabrit1986,cabrit1990}.

In the second case, since it is expected that more massive objects drive faster jets 
\footnote{The outflow velocity is related to the escape speed of the driving object, thus 
it increases as the square root of the mass of the central object \citep[e.g.][]{banerjee2006}.}, 
we would expect A1 to be more massive than A2. Indeed, A1 represents the source driving the faster jet. 
Moreover, the detection of brighter continuum emission associated with A1 (see Sect.~\ref{subsec:continuum})
may be interpreted as a larger envelope mass and, hence perhaps, larger protostellar mass.
However, this scenario 
is not supported by our WideX spectra at the positions of the detected continuum sources 
(see Fig.~\ref{fig:widex_A123}). 
In fact, a chemically rich spectrum, with emission from several COMs, is detected only towards A2, 
highlighting a hot-corino chemistry for this protostar. 

In the third case, 
A1 is expected to be younger than A2, as already suggested by \citet{choi2007,choi2010,choi2011}
based on the disks and outflows properties from VLA NH$_3$ and SiO observations. 
This view is supported by the detection of a hot-corino chemistry only towards A2 
(see Sect.~\ref{subsubsec:hotcorino} and Fig.~\ref{fig:widex_A123}).
Along with a slower 
jet driven by A2, associated with weaker H$_2$ shocked emission, a different chemistry  
may indicate a later evolutionary stage for this source. 
Indeed the lack of a detectable hot corino at A1 suggests a lower internal luminosity than at A2; 
in particular, the region where the dust temperature is high enough ($\geq$100~K) for the icy mantles to 
sublimate should be much smaller and more heavily beam diluted than at A2. This would imply 
a smaller stellar mass and/or earlier evolutionary stage for A1.
Further support for this view is given by the short dynamical timescale of the A1 jet.
In fact, for the farthest SiO knot along the A1 jet, assuming an observed velocity of 10$-$20~km~s$^{-1}$ (see Fig.~\ref{fig:spettri_A123}),
we can estimate a dynamical age of $\lesssim$10$^3$~yr. 
The A1 jet indeed appears as an intrinsically very young outflow.
On the other hand, according to \citet[][]{awad2010}, the chemically rich emission detected at A2 suggests 
an age of $\gtrsim$10$^4$$-$10$^5$ yr for this source, which is in agreement with the collapse age 
of 5$\times$10$^4$~yr derived by \citet{choi2010} from the rotation kinematics of the disk in NH$_3$.

%

\section{Summary and conclusions}
\label{sec:conclusions}

Our new PdBI observations of the IRAS4A binary system finally allowed us 
to clearly disentangle the jet emission from the Class 0 driving protostars, 
revealing the intrinsic different properties of the two jets.   
The detection of such a tight and collimated protobinary system is  
challenging for current theoretical models of multiple system formation.
We propose that the observed properties of the jets can be explained with 
different evolutionary stages for the two driving protostars.
In particular, the younger source (A1) drives a fast collimated jet associated with bright H$_2$ emission, 
whereas its sibling protostar (A2) powers a slower precessing jet 
and is associated with a rich hot-corino chemistry.   
However, a difference in the protostellar masses may not be excluded with certainty.
Indeed, information on the individual luminosities of the protostars would be crucial in order to 
confirm the proposed scenario.
Finally, jet variability, in particular episodic ejection \citep[see e.g.][]{codella2014}, may also play a role.

\begin{acknowledgements}
We thank Serena Viti and Brunella Nisini for fruitful discussions 
and helpful inputs on this project.
We are very grateful to the IRAM staff, whose dedication allowed
us to carry out the CALYPSO project. 
This work was partly supported by the ASI--INAF project 01/005/11/0, 
the PRIN INAF 2012 -- JEDI, and the Italian Ministero dell'Istruzione, 
Universit\`a e Ricerca through the grant Progetti Premiali 2012 -- iALMA.
\end{acknowledgements}


\Online

\begin{appendix} 
\section{Comparison with large-scale emission}
\label{Asec:SD}
In this section, a comparison between the IRAM-PdBI CO(2$-$1) data presented in this paper 
and the large-scale emission is presented. In particular, VLA SiO emission data from \citet{choi2005} and 
single-dish CO data by \citet{yildiz2012} are shown. 
The comparison shows the change of propagation direction 
from north-south to about $45^{\circ}$ in the large-scale CO emission 
\citep[see also][]{blake1995, girart1999,choi2001,choi2005}, which also seems to be consistent with the 
large-scale SiO data.
Our possible evidence of precession in the A2 jet on very small scales
(Fig.~\ref{fig:channelmaps}) is consistent with the large-scale 
CO and SiO morphologies, confirming that the A2 jet dominates the large-scale emission. 
Moreover, the non-detection in the IRAM-PdBI data of CO(2$-$1) emission associated with the A2 jet at distances larger than about 3$^{\prime\prime}$ 
from the source is possibly due to filtering of large-scale emission by the interferometer, suggesting that the CO emission 
associated with the A2 jet is quite extended.

\begin{figure}[h!]
\centering
\includegraphics[width=0.38\textwidth]{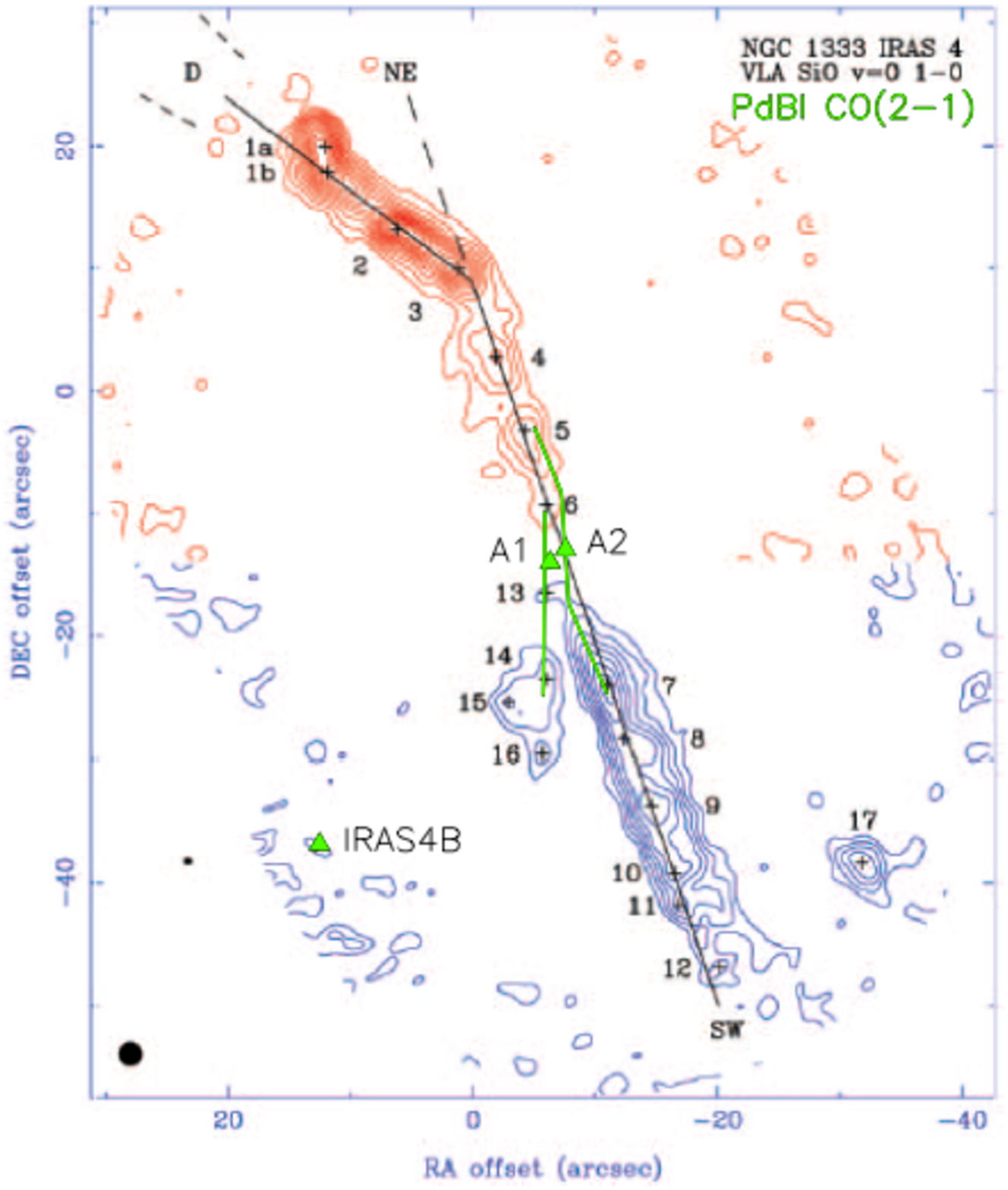}
\includegraphics[width=0.41\textwidth]{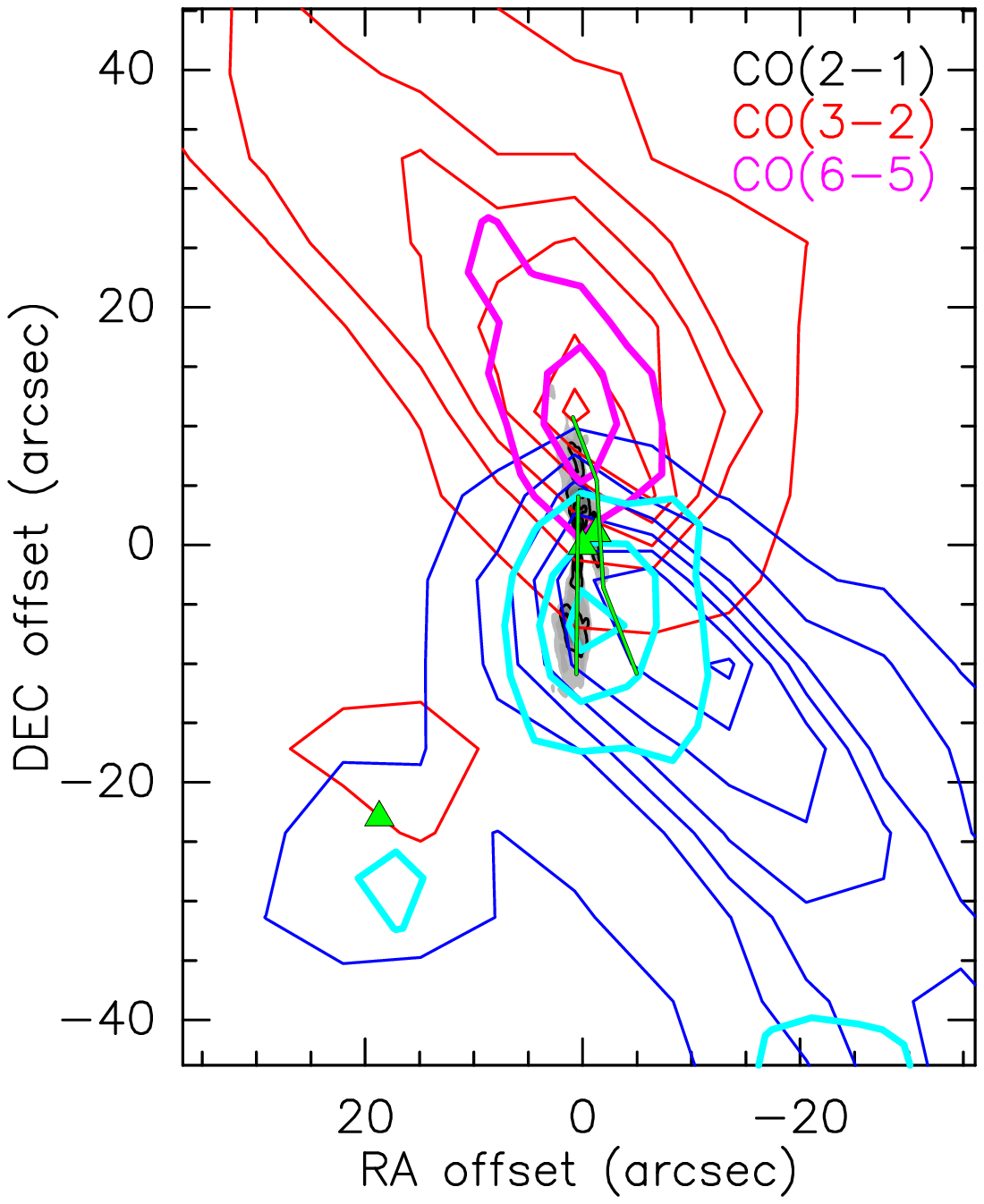}
\caption{\emph{Upper:} The VLA SiO(1$-$0) map (red and blue contours) by \citet{choi2005} is shown in comparison with the proposed propagation directions 
of the A1 and A2 jets (green solid lines, see Fig.~\ref{fig:channelmaps}). 
The positions of IRAS4A1 and A2 (this work) and of IRAS4B \citep{looney2000} are marked with green triangles. 
\emph{Lower:} The PdBI CO(2$-$1) map (grey scale and black contours) integrated over the whole velocity range of emission (-20, +55~km~s$^{-1}$) 
is compared with the JCMT CO(3$-$2) (red and blue contours) and APEX CO(6$-$5) (magenta and cyan contours) 
maps from \citet{yildiz2012}. CO(2$-$1) emission is only shown within the PdBI field of view of 20$^{\prime\prime}$ at 230~GHz.
The CO(3$-$2) and CO(6$-$5) maps are integrated in the velocity ranges between -20~km~s$^{-1}$ and 3~km~s$^{-1}$ 
for the blue-shifted emission and 12 km~s$^{-1}$ and 50~km~s$^{-1}$ for the red-shifted emission. 
The contour levels start at the 3~$\sigma$ level and increase in steps of 3~$\sigma$ for the CO(2$-$1) emission, 
from the 5~$\sigma$ level emission in steps of 10~$\sigma$ for the CO(3$-$2), 
and from the 5~$\sigma$ level emission in steps of 5~$\sigma$ for the CO(6$-$5).
}
\label{Afig:SD}
\end{figure}

\section{WideX spectra at the EHV peak}
\label{Asec:widex}
\begin{figure*}
\centering
\includegraphics[width=0.8\textwidth]{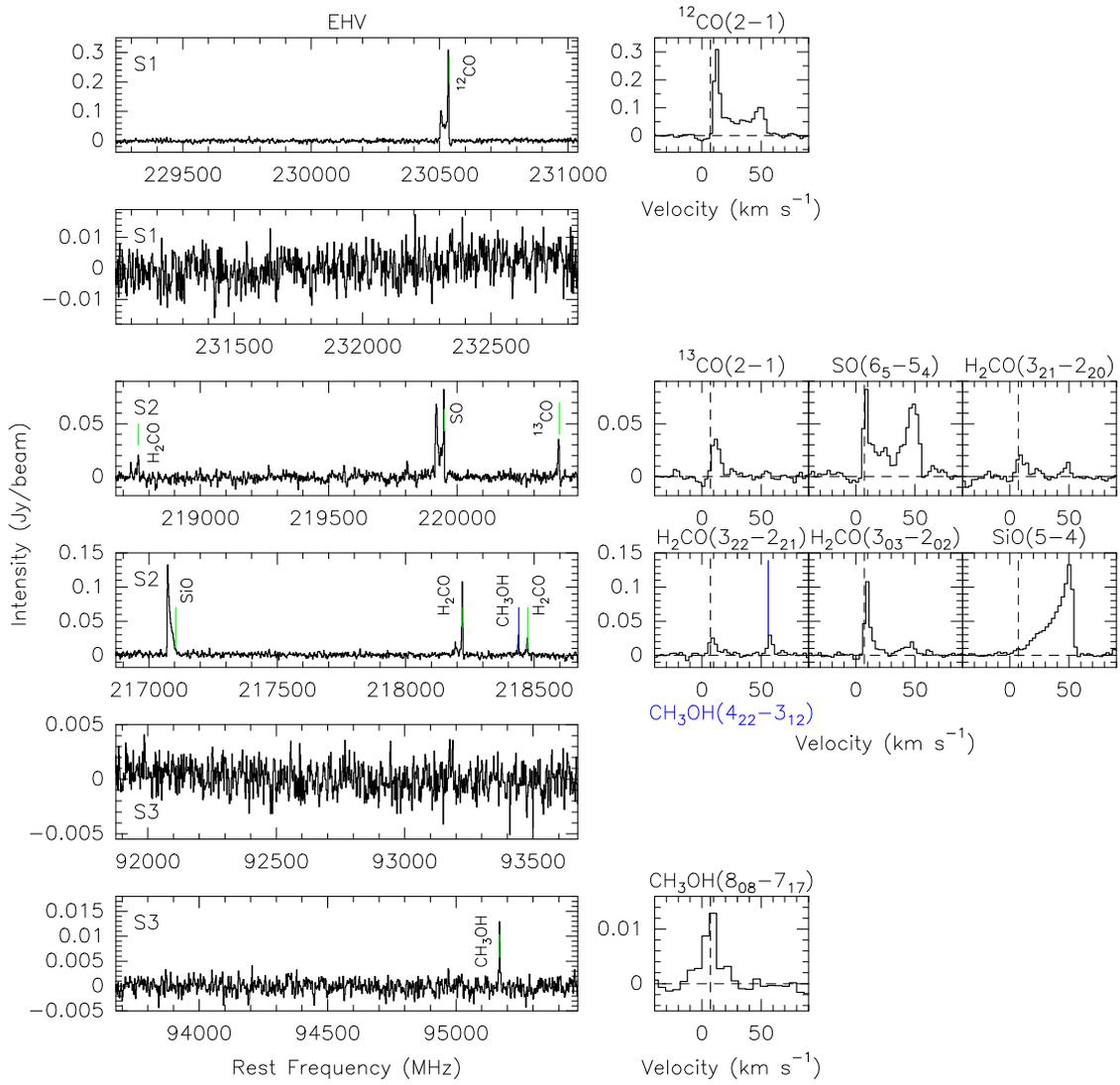}
\caption{Additional lines detected in the WideX spectra at the peak of the EHV gas.
}
\label{Afig:widexEHV}
\end{figure*}

Figure~\ref{Afig:widexEHV} shows the additional lines detected in the WideX spectra at the peak of the EHV gas that are analysed 
and discussed in the main text (see also Fig.~\ref{fig:spettri_EHV} and Sect.~\ref{subsec:EHVgas}). 
In particular, the CO(2$-$1), SO(6$_5$$-$5$_4$), H$_2$CO(3$_{21}$$-$2$_{20}$), and H$_2$CO(3$_{03}$$-$2$_{02}$) lines 
show a double-peaked profile 
with a velocity component peaking around the systemic velocity and a secondary velocity component around 50~km~s$^{-1}$;  the
SiO(5$-$4) line profile presents a broad wing emission that peaks at 50~km~s$^{-1}$ and extends down to the systemic velocity; 
finally, the $^{13}$CO(2$-$1), H$_2$CO(3$_{22}$$-$2$_{21}$), CH$_3$OH(4$_{22}$$-$3$_{12}$), 
and CH$_3$OH(8$_{08}$$-$7$_{17}$) line profiles 
show a single emission peak around systemic velocity, possibly due to a lack of sensitivity.

\end{appendix}

\end{document}